%% file: main.tex
\documentclass[sigconf,nonacm]{acmart}

\usepackage[table]{xcolor}
\definecolor{bgray}{rgb}{0.9,0.9,0.9}
\definecolor{impl}{HTML}{E0EEF0}
\definecolor{impline}{HTML}{3B5A8C}

\usepackage{xspace}
\usepackage{multirow}
\usepackage{tabularx}
\usepackage{threeparttable}
\usepackage{pifont}

\usepackage{tikz}
\usetikzlibrary{arrows.meta,positioning,calc}

\newcommand{\ourpackage}{\textsf{dqmeasure}\xspace}
\newcommand{\taberr}{\emph{tab\_err}\xspace}

\theoremstyle{acmdefinition} %
\newtheorem{definition}{Definition}

\newcommand{\dqm}[2][]{\textsc{\MakeLowercase{#2}}\if\relax\detokenize{#1}\relax\else~(#1)\fi}

\AtBeginMaketitle{%

\input{edbt-macros}
}%

\begin{document}

\title{Implementation Guidelines for Data Quality Metrics}

\newcommand{\bht}{\affiliation{%
  \institution{Berliner Hochschule für Technik}
  \city{Berlin}\country{Germany}}}
\newcommand{\ptb}{\affiliation{%
  \institution{Physikalisch-Technische Bundesanstalt}
  \city{Berlin}\country{Germany}}}
\newcommand{\ecdf}{\affiliation{%
  \institution{Einstein Center Digital Future}
  \city{Berlin}\country{Germany}}}
\newcommand{\hpi}{\affiliation{%
  \institution{Hasso Plattner Institute}
  \city{Potsdam}\country{Germany}}}
\newcommand{\fernuni}{\affiliation{%
  \institution{FernUniversität in Hagen}
  \city{Hagen}\country{Germany}}}
\newcommand{\jku}{\affiliation{%
  \institution{Johannes Kepler University Linz}
  \city{Linz}\country{Austria}}}

\author{Philipp Jung}
\bht
\orcid{0009-0003-2275-3038}
\email{philipp.jung@bht-berlin.de}

\author{Katinka Becker}
\ptb
\orcid{0009-0007-9663-6722}
\email{katinka.becker@ptb.de}

\author{Felix Biessmann}
\bht\ecdf   %
\orcid{0000-0002-3422-1026}
\email{felix.biessmann@bht-berlin.de}

\author{Valerie Restat}
\fernuni
\orcid{0000-0002-5960-5886}
\email{valerie.restat@fernuni-hagen.de}

\author{Martin Seyferth}
\ptb
\orcid{0009-0000-5930-287X}
\email{martin.seyferth@ptb.de}

\author{Daniel Schwabe}
\ptb
\orcid{0000-0003-2825-3352}
\email{daniel.schwabe@ptb.de}

\author{Lisa Ehrlinger}
\jku\hpi 
\orcid{0000-0001-5313-0368}
\email{lisa.ehrlinger@hpi.de}

\newcommand{\numDQMs}{20}
\newcommand{\numDSs}{3}

\renewcommand{\shortauthors}{Jung et al.} %

\begin{abstract}
\input{sections/0-abstract}
\end{abstract}


\maketitle

\input{sections/1-intro}
\input{sections/2-related-work}
\input{sections/3-executable-metrics}
\input{sections/4_implementation}
\input{sections/5_evaluation}
\input{sections/6_discussion}
\input{sections/7_conclusion}
\input{sections/8_artifacts}

\balance
\bibliographystyle{ACM-Reference-Format}
\bibliography{references}

\end{document}

%% file: edbt-macros.tex
\def\EDBTISSN{xxxx-xxxx}
\def\EDBTISBN{xxx-x-xxxxx-xxx-x}
\setcopyright{none}
\copyrightyear{\copyright\ 2027 Copyright held by the owner/author(s).
  Published on xxxxx under ISBN \EDBTISBN, series ISSN \EDBTISSN.
  Distribution of this paper is permitted under the terms of the
  Creative Commons license CC-by-nc-nd 4.0}
\acmYear{2027}
\acmDOI{}
\acmISBN{}

\acmConference[EDBT '27]{Extending Database Technology}{6-9 April 2027}{Lille (France)}

\settopmatter{printacmref=false, printccs=false, printfolios=false}

\newsavebox{\ximagebox}
\newlength{\ximageheight}
\newsavebox{\xglyphbox}
\newlength{\xglyphheight}
\newcommand{\xbox}[1]%
  {\savebox{\ximagebox}{#1}%
  \settoheight{\ximageheight}{\usebox{\ximagebox}}%
  \savebox{\xglyphbox}{\color{white}\char32}%
  \settoheight{\xglyphheight}{\usebox{\xglyphbox}}%
  \raisebox{\ximageheight}[0pt][0pt]{\raisebox{-\xglyphheight}[0pt][0pt]{%
    \makebox[0pt][l]{\usebox{\xglyphbox}}}}%
    \usebox{\ximagebox}%
    \raisebox{0pt}[0pt][0pt]{\makebox[0pt][r]{\usebox{\xglyphbox}}}}

\newsavebox{\LogoBox}
\sbox{\LogoBox}{}

\fancypagestyle{firstpagestyle}{%
  \fancyhf{}%
  \fancyfoot{}%
  \fancyhead[L,C]{}
  \fancyhead[R]{}{\raisebox{15pt}[0pt][0pt]{\xbox{\usebox{\LogoBox}}}}
  }

\hypersetup{%
  pdfcopyright={CC-by-nc-nd},
  pdfcreator={LaTeX with acmart, hyperref, and EDBT modifications},
  pdfvolumenum={XX},
  pdfissuenum={X},
  pdfisbn={\EDBTISBN},
  pdfissn={\EDBTISSN}
}

%% file: sections/0-abstract.tex
Despite decades of data quality (DQ) research, a gap remains between DQ dimensions, such as accuracy or completeness, which the literature defines in textual form, and DQ tools, which typically implement low-level checks that are not aligned with these dimensions. 
ISO/IEC 25024 and ISO/IEC 5259 attempt to bridge this gap by defining DQ metrics for each dimension. However, these DQ metrics are hardly used, 
because the standards leave open how to implement them: for example, the metric for syntactic accuracy counts syntactically accurate values, but does not state how to decide that a value is syntactically accurate.
This simply moves the problem to another level without solving it. As a result, DQ assessment currently cannot build on the standards.

In this paper, we make the ISO DQ metrics executable. We classify all data-level metrics of both standards into (i)~\emph{generalizable} metrics that need no input beyond the data, (ii)~\emph{parameterized} metrics whose parameters can be learned from clean reference data or set by an expert, and (iii)~\emph{non-generalizable} metrics that need qualitative judgment and cannot be automated.
For the metrics that can be automated, i.e., categories (i) and (ii), we propose implementation guidelines that resolve what the standards leave open. 
We realize the guidelines in \ourpackage, an open-source library of \numDQMs{} metrics that learns these parameters from reference data instead of relying on manually defined rules. 
Our experiments on real-world and synthetic datasets show that the metric scores decrease monotonically with an increasing number of injected errors, decline together with downstream ML performance, and scale linearly with the number of rows, which enables automated DQ monitoring based on the standards.

%% file: sections/1-intro.tex
\section{Introduction}
\label{sec:intro}

\input{figures/overview-figure-v2}

Data quality (DQ) describes the fitness of data for use, commonly characterized by different dimensions, such as accuracy, completeness, consistency, and timeliness~\cite{wang_1996,Redman_1996,Batini_2016}.
DQ dimensions can be quantified by one or more DQ \emph{metrics} (DQMs), assigning them a numeric score~\cite{Heinrich_2018,pipino_2002}.
With the increased adoption of machine learning (ML) components in modern software systems and the shift from model-centric to data-centric artificial intelligence (AI), DQ has become a central concern for enterprises~\cite{Zha_2025}. Since enterprise data is often distributed, heterogeneous, and stored in different formats~\cite{ehrlinger_2025,Bodensohn_2025}, detecting and cleaning errors remains a largely manual effort that takes time away from designing and training models~\cite{Abedjan_2016,Restat_2025}.
Poor DQ also affects downstream tasks, in particular ML pipelines~\cite{mohammed_2025,Li_2021}, reducing trust of users and causing harmful results~\cite{schelterFairPrepPromotingData2020,Zech_2018, Daneshjou_2022}. Moreover, regulation increases the pressure to assess DQ systematically: Article~10 of the EU AI Act requires that the training, validation, and test data of high-risk AI systems are relevant, sufficiently representative, and, to the best extent possible, free of errors and complete~\cite{eu_2024}.

Despite this pressure, a gap separates research and practice (cf. \autoref{fig:overview}).
The information systems community typically defines DQ dimensions in textual form without detailing their measurement, while DQ tools implement low-level checks that are not always aligned with these dimensions~\cite{papastergios_2026,ehrlinger_2022}.
To bridge this gap, ISO/IEC 25024~\cite{iso25024} made a first attempt to specify DQMs for each dimension, aligned with the DQ characteristics defined in ISO/IEC 25012~\cite{iso_2008}.
Due to the rise of ML, this was complemented in 2024 by ISO/IEC 5259~\cite{iso5259}, which defines DQMs scoped to analytics and ML.
Note that ISO standards use the term DQ \emph{measure} for what the DQ literature calls a \emph{metric}~\cite{Heinrich_2018}. We follow the literature and use metric throughout this work, including when referring to the standards.
Prior work builds on these standards conceptually, for example for DQ evaluation and certification frameworks~\cite{gualo_2021, merino_2016, gualo_2023}.

However, the ISO standards are hardly used in practice, since it is often unclear how their DQMs should be implemented~\cite{papastergios_2026}.
For example, ISO/IEC 25024 defines the DQM for syntactic accuracy, identified as Acc-I-1 in the standard, as the ratio~\cite{iso25024}
{\footnotesize
\[
\frac{\text{number of data items with syntactically accurate values}}{\text{number of data items for which syntactic accuracy is required}}.
\]}
Yet the standard does not explain how to determine that a data item has syntactically accurate values.
Hence, practitioners cannot directly compute standardized DQMs on their data, and tool-based measurements cannot be mapped back to the standards.

This raises our central research challenge: making the DQMs of ISO/IEC 25024 and ISO/IEC 5259 automatically applicable to real-world datasets. Automation is a prerequisite for continuous DQ monitoring in enterprise AI settings, where manual assessment does not scale~\cite{ehrlinger_2025,Ehrlinger_2017,Redyuk_2021}. 
\autoref{fig:overview} illustrates our approach: implementation guidelines that make the ISO DQMs executable and that are realized in \ourpackage{}.
In summary, we make the following contributions:
\begin{itemize}
    \item A model of executable DQMs and a new classification scheme that distinguishes which DQMs of ISO/IEC 25024 and ISO/IEC 5259 can be automated and which require external input, such as parameters or domain knowledge.
    \item Five implementation guidelines that describe how a DQM that is defined in textual form can be turned into an executable DQM.
    \item \ourpackage{}, a reference implementation of \numDQMs{} DQMs that follows these guidelines. 
    Inspired by established ML APIs, \ourpackage learns the parameters of a DQM from clean reference data instead of relying on manually defined rules, while experts can still set them. The library includes a DQM for semantic accuracy that is based on a large language model (LLM), and can be used in continuous integration pipelines and  for continuous DQ monitoring.
    \item An evaluation on real-world and synthetic datasets, which shows that the DQMs (i)~detect injected errors, (ii)~decline in score together with downstream ML performance, and (iii)~scale linearly with the number of rows.
\end{itemize}

The remainder of this paper is structured as follows.
Section~\ref{sec:related-work} reviews related work.
Section~\ref{sec:executable-metrics} introduces our model of executable DQMs, the classification scheme, and the implementation guidelines,  and Section~\ref{sec:implementation} describes their implementation in \ourpackage. 
We evaluate \ourpackage in Section~\ref{sec:eval}, discuss open challenges in Section~\ref{sec:discussion},  and conclude in Section~\ref{sec:conclusion}.

%% file: figures/overview-figure-v2.tex
\begin{figure}[t]
  \centering
  \makebox[\linewidth][c]{%
  \begin{tikzpicture}[
    font=\sffamily\small,
    >=Latex,
    node distance=2mm and 10mm
  ]

  \tikzset{
    block/.style={
      draw,
      thick,
      rounded corners=3pt,
      align=center,
      inner xsep=1.5pt,
      inner ysep=2pt,
      minimum height=15mm,
      text width=0.42\linewidth
    },
    contrib/.style={
      block,
      thick,
      draw=impline,
      fill=impl
    },
    arrow/.style={->, semithick, impline}
  }

  \node[block, anchor=north west] (high) at (0,0) {%
    \textbf{DQ dimensions}\\
    {\footnotesize\itshape (standard-defined)}\\
    {\footnotesize\textit{e.g.,} accuracy, completeness, \ldots}
  };

  \node[block, anchor=north east] (low) at (\linewidth,0) {%
    \textbf{Low-level checks}\\
    {\footnotesize\itshape (tool-specific)}\\
    {\footnotesize\textit{e.g.,} missing values, ranges, \ldots}
  };

  \coordinate (contribcenter) at ([yshift=-26mm]$(low)!0.5!(high)$);
  \coordinate (contriby) at (contribcenter);
  \coordinate (contribcenter) at ([xshift=\dimexpr0.32\linewidth+6mm\relax]high.west |- contriby);

  \node[contrib, anchor=south east, xshift=10mm, text width=0.35\linewidth, minimum height=10mm] (guidelines)
    at (contribcenter) {%
      \textbf{Implementation\\ guidelines}
    };

  \node[contrib, anchor=south east, text width=32mm, minimum height=12mm] (pkg)
    at (low.east |- contriby) {%
      \textbf{\ourpackage}\\
      {\footnotesize\texttt{SyntacticDataAccuracy(c)}\\
      \texttt{.fit(reference)}
      \texttt{.score(data)}\ $\rightarrow$\ 0.93}
    };

  \draw[<->, semithick, dashed, red!45!black, shorten <=2pt, shorten >=2pt]
    (high.east) -- node[above=4pt, fill=white, inner sep=1pt, align=center] {\footnotesize\itshape gap} (low.west);
    
  \coordinate (dim2guideStart) at ($(guidelines.west |- high.south)+(-3mm,0)$);
  \draw[arrow] (dim2guideStart) |- (guidelines.west);
  \draw[arrow] (guidelines.east) -- (guidelines.east -| pkg.west);

  \node[align=center, anchor=north, fill=white, inner sep=2pt, rounded corners=1pt, font=\footnotesize\rmfamily]
    (highmetric) at ([yshift=-1mm]high.south)
    {\textbf{Acc-I-1}\\Syntactic data accuracy};

  \draw[arrow] ([xshift=0mm]pkg.north) -- ([xshift=0mm]pkg.north |- low.south);

  \end{tikzpicture}%
  }

  \caption{Gap between standard-defined DQ dimensions and tool-specific low-level checks: for Acc-I-1, ISO/IEC 25024 defines the metric but leaves open how to decide that a value is ``syntactically accurate''. Our contribution (blue) bridges this gap with implementation guidelines that are realized in the open-source library \textit{\ourpackage{}}.}
  \Description{Figure showing an operationalization gap between standard-defined data quality dimensions and tool-specific low-level checks. Two top boxes labeled DQ dimensions and Low-level checks are separated by a dashed double-headed arrow labeled "gap". Two blue boxes below highlight the contribution: Implementation guidelines and the dqmeasure library, connected by blue arrows.}
  \label{fig:overview}
\end{figure}
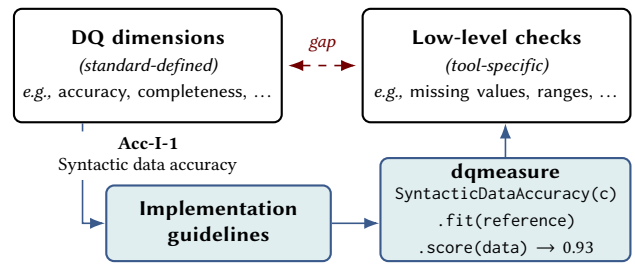

%% file: sections/2-related-work.tex
\section{Related Work}
\label{sec:related-work}

To embed our work into existing literature, we first review research on DQ dimensions and metrics (\autoref{sec:dqm_definitions}), then DQ tools that are used in practice (\autoref{sec:dqm_implementations}), and finally work that bridges the gap between ISO standards and their implementation (\autoref{sec:rel-work-bridge}).

\subsection{Defining Data Quality Dimensions}
\label{sec:dqm_definitions}
Wang and Strong~\cite{wang_1996} established DQ as a multidimensional concept: from a survey of data consumers, they derived an early classification of 15 DQ dimensions into four categories (intrinsic, contextual, representational, and accessibility).
Since then, more than 200 dimensions have been proposed~\cite{Berti_2007}, along with several classifications~\cite{Scannapieco_2002,Laranjeiro_2015, cichy_2019}, such that the definition and interpretation of a dimension differ considerably between sources~\cite{ehrlinger_2025}.
Other studies propose specific DQ frameworks or weighted metric-based approaches~\cite{vaziri_2019, elouataoui_2022, schwabe_2024} or examine specific DQ dimensions and metrics in more detail~\cite{heinrich_2009, haegemans_2016, ehrlinger_2018}. 
ISO/IEC 25012~\cite{iso_2008} standardizes 15 DQ characteristics, five of which describe properties of the data itself (inherent point of view); in this paper, we focus on four of these dimensions, plus timeliness~\cite{iso5259}:
\begin{itemize}
    \item \textbf{Accuracy}: the degree to which data correctly represents the true value of the intended attribute of a concept or event.
    \item \textbf{Completeness}: the degree to which data has values for all expected attributes and related entity instances.
    \item \textbf{Consistency}: the degree to which data is free from contradiction and coherent with other data.
    \item \textbf{Currentness}: the degree to which data is of the right age.
    \item \textbf{Timeliness}: the degree to which data becomes available in time after the phenomenon it describes occurred.
\end{itemize}

\textbf{From dimensions to metrics.}
DQ dimensions describe the characteristics of high-quality data, but do not provide quantitative measures~\cite{Wand_1996, Batini_2016}. 
A DQ metric is a function that quantifies a DQ dimension with a numerical score, typically within the interval $[0,1]$, and indicates how well a dataset satisfies that DQ dimension~\cite{Heinrich_2018,Wand_1996}.
Heinrich et al.~\cite{Heinrich_2018} propose five requirements for such metrics, building on Pipino et al.~\cite{pipino_2002}: the existence of minimum and maximum values, interval-scaled values, the quality of configuration parameters, sound aggregation, and economic efficiency. 
Note that these requirements do not define new dimensions, but constrain how a metric should behave as a measurement function, independent of the dimension it quantifies. Although they are useful for defining metrics, their practical interpretation remains challenging: Bronselaer et al.~\cite{bronselaer_2018} argue that even normalized and interval-scaled metrics are not always easy to interpret, and Laranjeiro et al.~\cite{Laranjeiro_2015} show that the mapping between data errors and dimensions is not always straightforward.
Furthermore, metrics quantify DQ at different levels, i.e., data, schema, or metadata, depending on which object they target~\cite{Batini_2016}. In this paper, we focus on data-level metrics.

\subsection{Data Quality Metrics Implementations}
\label{sec:dqm_implementations}
In practice, DQ is assessed with tools that allow users to define and execute data validation checks~\cite{ehrlinger_2022, seyferth_2026}. Examples include \texttt{deequ}~\cite{schelter_2018}, \texttt{dbt Core}~\cite{dbtcore_2026}, \texttt{Evidently AI}~\cite{evidentlyai_2026}, \texttt{Great Expectations}~\cite{greatexpectations_2026}, \texttt{pandera}~\cite{pandera_2026}, and \texttt{pydantic}~\cite{pydantic_2026}. 
These tools are widely adopted: each of them has several thousand to more than twenty thousand stars on GitHub\footnote{As of August 2026.}, which indicates that a large number of practitioners implement DQ checks on a daily basis. However, the tools do not tell their users what to implement. Several surveys~\cite{ehrlinger_2022, papastergios_2026} show that the tools provide low-level checks, e.g., for missing values or value ranges, and leave it to the user to specify the rules and parameters, while generally applicable DQ metrics are hardly implemented~\cite{ehrlinger_2022}. Moreover, the checks are not aligned with DQ dimensions: Papastergios et al.~\cite{papastergios_2026} reveal many-to-many relationships between the functionalities of the tools and the dimensions of ISO/IEC 25012. As a consequence, practitioners implement their metrics ad hoc, and the results can neither be compared across tools nor mapped back to a DQ dimension or a standard.

Our implementation guidelines address this gap by describing how DQ metrics defined in a standard can be implemented systematically and independent of a specific tool. 
\ourpackage is a reference implementation that follows these guidelines. For its evaluation, we build on work that injects errors into datasets and measures their impact on downstream ML tasks~\cite{schelterJENGAFrameworkStudy2021,jung2025realistic}.

\subsection{Bridging Standards and Implementations}
\label{sec:rel-work-bridge}

Several works address how DQ metrics are defined and how their measurement is organized. Pipino et al.~\cite{pipino_2002} introduce functional forms for DQ metrics, such as the simple ratio, and Bronselaer et al.~\cite{bronselaer_2018} provide a measure-theoretic foundation for DQ based on predicates. With respect to standardization, ISO/IEC 25021~\cite{iso25021} specifies how quality measures are defined based on quality measure elements, which is the structure that ISO/IEC 25024 and ISO/IEC 5259 follow. Caballero et al.~\cite{Caballero_2007} propose a DQ measurement information model based on ISO/IEC 15939 to unify the terminology of the measurement process, complemented by a methodology to draw up DQ measurement processes in organizations~\cite{Caballero_2008}. These works describe how DQ metrics are defined, but not how they are implemented.

The practical application of ISO/IEC 25024 has been examined in a line of work that evaluates the DQ metrics by means of business rules~\cite{gualo_2021,caballero_2022,galera_2023}. Gualo et al.~\cite{gualo_2021} report on their experiences with an assessment process based on ISO/IEC 25012 and ISO/IEC 25024 in three organizations and conclude that DQ assessment and certification is beneficial for them. A central part of their process is the identification of business rules, which turned out to be the most time-consuming step. Caballero et al.~\cite{caballero_2022} therefore propose a methodology for systematically collecting, grouping, and validating business rules, which Galera et al.~\cite{galera_2023} extend to support the identification and classification of business rules and their mapping to the DQ metrics of the standards. Nevertheless, the creation of business rules remains a manual and time-consuming step. This is in line with other work that highlights the manual effort involved in defining DQ tests~\cite{schelter_2018}, which hinders the automation of DQ assessment~\cite{biessmannAutomatedDataValidation}.

In contrast, we focus on the implementation of DQ metrics that are already defined in ISO/IEC 25024 and ISO/IEC 5259. Instead of manually defining a set of rules for each metric, we determine which metrics can be automated, learn the required parameters from reference data whenever possible, and propose implementation guidelines that describe the remaining steps.

%% file: sections/3-executable-metrics.tex
\section{From Definitions to Executable Metrics}
\label{sec:executable-metrics}

We first introduce a simple model that describes what is needed to execute a DQM (\autoref{sec:model}). Based on this model, we discuss the inputs that DQMs require and classify the DQMs of the ISO standards according to their generalizability (\autoref{sec:dqm-classification}). Finally, we propose implementation guidelines, which describe step by step how a textual metric definition can be turned into an executable metric (\autoref{sec:guidelines}).

\subsection{A Model of Executable DQMs}
\label{sec:model}

We consider a dataset $D$ as a table with a set of rows $R$ and a set of columns $C$. A cell $(r, c) \in R \times C$ holds a data value or \textsc{NULL}. \autoref{tab:iso_concepts_to_dqmeasure} shows how the terms used in the ISO standards map to this notation.
Note that the standards do not use the term \emph{data item} consistently: depending on the DQM, it can refer to a column or a cell. While \autoref{tab:iso_concepts_to_dqmeasure} shows the default mapping, the elements a DQM counts are specified by its units (see Definition~\ref{def:dqm}).

\begin{table}[ht]
  \centering
  \caption{Mapping of ISO concepts onto \ourpackage{}.}
  \label{tab:iso_concepts_to_dqmeasure}
  \begin{tabular}{p{2.5cm} p{2.5cm} p{2cm}}
    \toprule
    \textbf{ISO/IEC 25024} & \textbf{ISO/IEC 5259} & \textbf{In this paper} \\
    \midrule
    data value & data value & cell value \\
    data item, field & data item, field & cell \\
    attribute & feature & column \\
    data record, tuple & data record & row \\
    data file, data set & data frame, dataset & table \\
    data type & data type & column dtype \\
    \bottomrule
  \end{tabular}
\end{table}

Most DQMs in the ISO standards~\cite{iso25024, iso5259} are defined as a ratio $X = A/B$, where $B$ is the number of elements for which a requirement applies, and $A$ is the number of those elements that fulfill the requirement, which corresponds to the simple ratio by Pipino et al.~\cite{pipino_2002}. 
In the terminology of ISO/IEC 25021~\cite{iso25021}, $X(A,B)$ is a \emph{measurement function} over two quality measure elements $A$ and $B$, each of which is obtained by a measurement method. 
However, the standards describe the measurement method of $A$ and $B$ only in textual form. We therefore do not introduce a new way of defining DQMs, but make this measurement method explicit by specifying four components, which we define as follows:

\begin{definition}[Executable DQM]
\label{def:dqm}
An executable DQM is a tuple $(s, U, \theta, \varphi)$, where
\begin{itemize}
  \item the \emph{subject} $s \in C \cup \{D\}$ is either a column or the entire table, for which one score is reported,
  \item the \emph{units} $U$ are the elements that are counted (e.g., the cells of a column or the rows of the table),
  \item the \emph{reference} $\theta$ comprises all parameters that are required to evaluate the DQM (e.g., an interval or a set of valid values), and
  \item the \emph{condition} $\varphi_\theta : U \to [0, 1]$ specifies for each unit to which degree it fulfills the requirement.
\end{itemize}
\end{definition}
The score of the DQM is calculated as
\begin{equation}
  \label{eq:dqm-score}
  X = \frac{A}{B} = \frac{\sum_{u \in U} \varphi_\theta(u)}{|U|}.
\end{equation}

For example, Data accuracy range (Acc-I-7) verifies whether the values of a column are within a required interval. Here, the subject is the column, the units are all cells of the column that are not \textsc{NULL}, the reference is the interval $\theta = [\mathit{low}, \mathit{high}]$, and the condition is $1$ if a cell's value is within the interval, and $0$ otherwise. 
In contrast, Record completeness (Com-ML-4) has the entire table as subject and the rows as units. The condition is $1$ if a row contains no \textsc{NULL}, and no reference is required ($\theta = \emptyset$).

Note that the units are not necessarily cells of the subject. For Value occurrence completeness (Com-ML-2), the units are the expected occurrences of each value in the domain, which have no position in the table. In addition, the condition may consider more information than the unit itself: to evaluate Semantic data accuracy (Acc-I-2) for a cell through functional dependencies (FDs), the remaining values in the same row are required. We denote this additional information as the \emph{context} of the condition.

In addition to Definition~\ref{def:dqm}, we use the following two conventions. First, all scores are oriented such that a higher value indicates better DQ. Note that this is not the case in the ISO standards: for some DQMs, typically denoted as ``Risk of \ldots'', a value of $0$ represents the best outcome, and we alter the measurement function to report $1 - X$ for these DQMs. Second, if no units exist ($U = \emptyset$), the score is undefined and not $0$. 
With these conventions, each DQM is normalized to $[0,1]$ and uniformly oriented to support a sound aggregation of scores, which is in line with the requirements for DQMs by Heinrich et al.~\cite{Heinrich_2018}.

We note that a few DQMs in the standards have measurement functions other than $X = A / B$ or $X = 1 - A / B$, such as Distribution of data values (Con-ML-2), which compares the distribution of a column against a reference distribution. These DQMs are not covered by Definition~\ref{def:dqm} and are discussed separately in \autoref{sec:implementation}.

\subsection{A Generalizability-Based Classification Scheme for DQMs}
\label{sec:dqm-classification}
While pursuing the goal of DQM automation, we note that not all DQMs can be implemented the same way. Some metrics require additional input or knowledge, while others depend solely on the data itself. 
Based on the data-level DQMs in the ISO standards \cite{iso25024,iso5259}, we identified three different types of input that may be required to compute these metrics:

\begin{itemize}
\item \emph{Reference data:} A clean dataset without data quality problems with expected distributions of data values (e.g. ground truth of the target population).
\item \emph{Expert knowledge:} A qualitative input given by an expert (e.g., the update frequency required for a type of data).
\item \emph{Method:} A computational method to estimate a parameter (e.g. an outlier detection method).
\end{itemize}

In \autoref{table:allDQMs}, we present all data-level DQMs from ISO/IEC 25024~\cite{iso25024} and ISO/IEC 5259~\cite{iso5259} together with their required inputs.
We added a mapping between both standards for two cases: where ISO~5259 refers to ISO~25024 directly (grey cells in  \autoref{table:allDQMs}), or where the DQMs of both standards are defined differently, but compute the same value under the assumptions in~\autoref{tab:iso_concepts_to_dqmeasure}.

In many cases, the three types of input are not mutually exclusive, and a DQM may require multiple inputs at once, e.g., reference data together with a method that estimates a parameter to compare to the reference. In other cases, different types of input can substitute for one another, so that providing one type of input makes another unnecessary, e.g., a parameter that is set by an expert does not need to be learned from reference data.

Depending on the input required, DQMs can either be generalized directly, not at all, or with the additional step of obtaining input parameters.
To make this distinction of DQMs explicit, we propose a new classification scheme for the DQMs defined in ISO/IEC 25024~\cite{iso25024} and ISO/IEC 5259~\cite{iso5259}, grouping them according to their generalizability.
We identified three categories (column ``Category'' in \autoref{table:allDQMs}), which we define as follows:

\paragraph{Generalizable (G)} A DQM is generalizable if it is a fixed canonical assertion that does not require a parameter for evaluation. 
Examples are Value completeness (Com-ML-1) or Feature completeness (Com-ML-3 and Com-I-2), which simply count missing values.
Category G DQMs can be implemented directly, assuming that all required information is available in the data. We note that this assumption might not always hold in practice. 
An example that illustrates this assumption is the DQM Audited records (Aud-ML-1), which can be measured by counting if this feature is available. If the audit information is not stored in the data, Aud-ML-1 requires additional input.

\paragraph{Generalizable with parameterized knowledge (K)} A DQM is of category K if it requires one or more parameters to evaluate, e.g., Risk of inaccuracy (Acc-ML-4 and Acc-I-4), for which the number of outliers is required. 
Category K DQMs can be implemented by fitting or profiling parameters from reference data, sometimes using heuristics, or set manually by an expert user.

\paragraph{Non-generalizable (N)} A DQM is of category N if it requires qualitative input to evaluate. 
An example is User accessibility (Acs-I-1) which depends on the subjective judgement of a specific user.
Category N DQMs cannot be automated, since they require qualitative input in terms of expert judgement or user opinion.

\medskip
In terms of Definition~\ref{def:dqm}, the three categories differ in the reference $\theta$: for category G, no reference is required ($\theta = \emptyset$); for category K, $\theta$ is fitted from reference data or set by an expert; and for category N, $\theta$ cannot be specified as a parameter.

Based on this categorization, we implement all DQMs of categories G and K that can be evaluated on a single table without information about the surrounding system, and that belong to the DQ dimensions accuracy, completeness, consistency, currentness and timeliness, resulting in \numDQMs{} DQMs (see \autoref{sec:implementation}).
We selected these dimensions because they describe properties of the data itself, which corresponds to the inherent point of view in ISO/IEC 25012~\cite{iso_2008}.
For category K (see \autoref{sec:implementation}), we provide the user with the option to fit the input parameter from reference data whenever possible, but leave the option to include expert knowledge instead if applicable.
Additionally, since our implementation aims for out-of-the-box usability, we opt for default methods (e.g. an outlier detection method) whenever reasonable to include as many DQMs as possible.

\input{tables/TableAllDQMs}

\subsection{Implementation Guidelines}
\label{sec:guidelines}
\input{figures/guidelines-v2}

Based on the model and the classification scheme, we propose five implementation guidelines (IG1--IG5), which are summarized in \autoref{fig:guidelines}. The guidelines describe the steps that are required to turn a textual definition of a DQM into an executable DQM. Although we derived them while implementing the ISO DQMs, they are not restricted to the ISO standards and can be applied to any DQM that is defined in textual form.

\paragraph{IG1: Specify} In the first step, the four components of Definition~\ref{def:dqm} need to be specified. The \emph{subject} determines whether the DQM yields a score for a single column or for the entire table. A DQM should return exactly one score per subject; to assess several columns, the DQM is applied to each column separately. The \emph{units} determine what is counted, e.g., cells for Syntactic data accuracy (Acc-I-1), rows for Record completeness (Com-ML-4), or expected occurrences for Value occurrence completeness (Com-ML-2), and thus how the score is to be interpreted. The \emph{condition} specifies when a unit fulfills the requirement and corresponds to what Caballero et al.~\cite{caballero_2022} denote as business rule. All information that is required to evaluate the condition and that is not available in the data itself belongs to the \emph{reference} $\theta$.

\paragraph{IG2: Categorize} In the second step, the DQM is assigned to one of the three categories from \autoref{sec:dqm-classification}, depending on its reference. The category determines how to proceed: DQMs of category G do not require any parameters and can be implemented directly, DQMs of category K need to be parameterized (IG3), and DQMs of category N cannot be automated.

\paragraph{IG3: Parameterize} In the third step, the parameters of a category K DQM are set. The reference should be learnable from reference data, while an expert user should always be able to overwrite the learned values. In some cases, a definition does not specify the method to be used, e.g., how outliers are detected or which distribution measure is appropriate. Here, we recommend to select one default method that does not require further parameters, and to expose the method as a parameter that can be changed by the user. In other words, a DQM should be executable without any manual configuration.

\paragraph{IG4: Harmonize} In the fourth step, the results of all DQMs are harmonized such that they can be compared and reported together. This includes the uniform orientation of the scores and the handling of an empty set of units, as described in \autoref{sec:model}. In addition, missing values need to be handled consistently: we consider a value as missing if and only if it is \textsc{NULL}, which requires that placeholders (e.g., ``?'' or $-999$) are replaced beforehand. Missing values are excluded from the units, unless the DQM itself measures them. If the units correspond to cells or rows of the table, the result of the condition should be returned for each unit in addition to the score, which allows a user to trace a low score back to the affected cells or rows.

\paragraph{IG5: Validate} Finally, it needs to be ensured that a DQM fulfills its intended purpose. In contrast to IG1--IG4, this step checks the DQM rather than creates it. On the one hand, it should be documented for which data types and which kind of columns a DQM is meaningful. For example, Risk of data inconsistency (Con-I-3) counts duplicate values and is therefore meaningful for identifiers, but not for categorical columns. On the other hand, the implementation should be tested by injecting errors into a clean dataset: the score of a DQM should decrease with an increasing number of errors that belong to its DQ dimension, and it should remain stable for errors of other DQ dimensions. We follow this procedure in our evaluation in \autoref{sec:eval}. Validation becomes particularly important when parts of the previous steps are automated, e.g., when references or conditions are learned from data or generated by large language models, since automatically generated rules (or in our case metrics) need to be filtered for correctness and generalizability before they can be used in practical settings~\cite{Glock_2026}.

%% file: tables/TableAllDQMs.tex
\begin{table*}[]
\centering
\caption{List of data-level DQMs from ISO/IEC 25024~\cite{iso25024} and ISO/IEC 5259~\cite{iso5259} with their generalizability category and required inputs. Dark grey cells mark direct references from ISO/IEC 5259 to ISO/IEC 25024. The remaining rows with two IDs compute the same value under the assumptions in~\autoref{tab:iso_concepts_to_dqmeasure}. Light blue rows mark the DQMs implemented in \ourpackage{}, which are all DQMs for accuracy, completeness, consistency, currentness and timeliness with category G or K.}
\resizebox{0.94\textwidth}{!}{%
\begin{tabular}{llll|l|ccc}
\toprule
\textbf{Dimension}          & \textbf{ID (5259)}                        & \textbf{ID (25024)}                      & \textbf{Name}                                    & \textbf{Category} & \textbf{Reference} & \textbf{Expert} & \textbf{Method} \\ \midrule
Accessibility      & \cellcolor[HTML]{C0C0C0}Acs-ML-1 & \cellcolor[HTML]{C0C0C0}Acs-I-1 & User accessibility                      & N        &           & x      &        \\ \hline
Accessibility      & Acs-ML-3                         &                                 & Data accessibility                      & G        &           &        &        \\ \hline
\rowcolor{impl} Accuracy           & \cellcolor[HTML]{C0C0C0}Acc-ML-1 & \cellcolor[HTML]{C0C0C0}Acc-I-1 & Syntactic data accuracy                 & K        & x         & x      &        \\ \hline
\rowcolor{impl}Accuracy           & \cellcolor[HTML]{C0C0C0}Acc-ML-2 & \cellcolor[HTML]{C0C0C0}Acc-I-2 & Semantic data accuracy                  & K        & x         &        &        \\ \hline
\rowcolor{impl}Accuracy           & \cellcolor[HTML]{C0C0C0}Acc-ML-4 & \cellcolor[HTML]{C0C0C0}Acc-I-4 & Risk of inaccuracy                      & K        & x         &        & x      \\ \hline
\rowcolor{impl}Accuracy           & \cellcolor[HTML]{C0C0C0}Acc-ML-6 & \cellcolor[HTML]{C0C0C0}Acc-I-7 & Data accuracy range                     & K        & x         & x      &        \\ \hline
Auditability       & Aud-ML-1                         &                                 & Audited records                         & G        &           &        &        \\ \hline
Auditability       & Aud-ML-2                         &                                 & Auditable records                       & G        &           &        &        \\ \hline
Availability       &                                  & Ava-D-1                         & Data availability ratio                 & G        &           &        &        \\ \hline
Availability       &                                  & Ava-D-2                         & Probability of data availability        & G        &           &        &        \\ \hline
Balance            & Bal-ML-1                         &                                 & Brightness (image)                      & G        &           &        &        \\ \hline
Balance            & Bal-ML-2                         &                                 & Resolution (image)                      & G        &           &        &        \\ \hline
Balance            & Bal-ML-3                         &                                 & Images between cat                      & G        &           &        &        \\ \hline
Balance            & Bal-ML-4                         &                                 & Bounding box (images)                   & G        &           &        &        \\ \hline
Balance            & Bal-ML-5                         &                                 & Category bounding box                   & G        &           &        &        \\ \hline
Balance            & Bal-ML-6                         &                                 & Sample bounding box                     & G        &           &        &        \\ \hline
\rowcolor{impl}Completeness       & Com-ML-1                         &                                 & Value completeness                      & G        &           &        &        \\ \hline
\rowcolor{impl}Completeness       & Com-ML-2                         &                                 & Value occurrence completeness           & K        & x         &        &        \\ \hline
\rowcolor{impl}Completeness       & Com-ML-3                         & Com-I-2                         & Feature completeness                    & G        &           &        &        \\ \hline
\rowcolor{impl}Completeness       & Com-ML-4                         &                                 & Record completeness                     & G        &           &        &        \\ \hline
\rowcolor{impl}Completeness       & Com-ML-5                         &                                 & Label completeness                      & G        &           &        &        \\ \hline
\rowcolor{impl}Completeness       &                                  & Com-I-1                         & Record completeness                     & G        &           &        &        \\ \hline
Completeness       &                                  & Com-I-3                         & Data file completeness                  & N        &           & x      &        \\ \hline
Completeness       &                                  & Com-I-4                         & Data value completeness                 & N        &           & x      &        \\ \hline
\rowcolor{impl}Completeness       &                                  & Com-I-5                         & Empty records in a data file            & G        &           &        &        \\ \hline
Compliance         & Cmp-ML-1                         & Cmp-I-1                         & Data item compliance                    & K        & x         &        &        \\ \hline
Compliance         &                                  & Cmp-D-1                         & Regulatory compliance due to technology & N        &           & x      &        \\ \hline
Confidentiality    &                                  & Cnf-I-1                         & Encryption usage                        & G        &           &        &        \\ \hline
\rowcolor{impl}Consistency        & Con-ML-1                         & Con-I-3                         & Data record consistency                 & G        &           &        &        \\ \hline
\rowcolor{impl}Consistency        & & Con-I-3                         & Risk of data inconsistency                 & G        &           &        &        \\ \hline
\rowcolor{impl}Consistency        & Con-ML-2                         &                                 & Distribution of data values             & K        &           &        & x      \\ \hline
\rowcolor{impl}Consistency        & \cellcolor[HTML]{C0C0C0}Con-ML-3 & \cellcolor[HTML]{C0C0C0}Con-I-2 & Data format consistency                 & K        & x         &        &        \\ \hline
\rowcolor{impl}Consistency        & \cellcolor[HTML]{C0C0C0}Con-ML-4 & \cellcolor[HTML]{C0C0C0}Con-I-6 & Semantic consistency                    & K        & x         &        & x      \\ \hline
Credibility        & \cellcolor[HTML]{C0C0C0}Cre-ML-2 & \cellcolor[HTML]{C0C0C0}Cre-I-2 & Source credibility                      & K        & x         & x      &        \\ \hline
\rowcolor{impl}Currentness        & Cur-ML-1                         &                                 & Feature currentness                     & K        &           & x      & x      \\ \hline
\rowcolor{impl}Currentness        & Cur-ML-2                         &                                 & Record currentness                      & K        &           & x      & x      \\ \hline
\rowcolor{impl}Currentness        &                                  & Cur-I-1                         & Update frequency                        & K        & x         & x      &        \\ \hline
\rowcolor{impl}Currentness        &                                  & Cur-I-2                         & Timeliness of update                    & K        & x         & x      &        \\ \hline
Diversity          & Div-ML-1                         &                                 & Label richness                          & G        &           &        &        \\ \hline
Diversity          & Div-ML-2                         &                                 & Relative label abundance                & G        &           &        &        \\ \hline
Diversity          & Div-ML-3                         &                                 & Component richness                      & N        &           & x      &        \\ \hline
Diversity          & Div-ML-4                         &                                 & Category size diversity                 & N        &           & x      &        \\ \hline
Effectiveness      & Eft-ML-1                         &                                 & Feature effectiveness                   & N        &           & x      &        \\ \hline
Effectiveness      & Eft-ML-2                         &                                 & Category size effectiveness             & N        &           & x      &        \\ \hline
Effectiveness      & Eft-ML-3                         &                                 & Label effectiveness                     & N        &           & x      &        \\ \hline
Efficiency         &                                  & Eff-I-1                         & Efficient data item format              & K        & x         &        &        \\ \hline
Efficiency         &                                  & Eff-I-2                         & Usable efficiency                       & N        &           & x      &        \\ \hline
Efficiency         & \cellcolor[HTML]{C0C0C0}Eff-ML-1 & \cellcolor[HTML]{C0C0C0}Eff-D-1 & Data format efficiency                  & K        & x         &        &        \\ \hline
Efficiency         & \cellcolor[HTML]{C0C0C0}Eff-ML-2 & \cellcolor[HTML]{C0C0C0}Eff-D-2 & Data process efficiency                 & N        &           & x      &        \\ \hline
Efficiency         &                                  & Eff-D-5                         & Time delay of data update               & G        &           &        &        \\ \hline
Identifiability    & Idn-ML-1                         &                                 & Identifiability ratio                   & N        &           & x      &        \\ \hline
Portability        & \cellcolor[HTML]{C0C0C0}Por-ML-1 & \cellcolor[HTML]{C0C0C0}Por-D-1 & Data portability ratio                  & G        &           &        &        \\ \hline
Portability        & \cellcolor[HTML]{C0C0C0}Por-ML-2 & \cellcolor[HTML]{C0C0C0}Por-D-2 & Prospective portability                 & G        &           &        &        \\ \hline
Precision          & \cellcolor[HTML]{C0C0C0}Pre-ML-1 & \cellcolor[HTML]{C0C0C0}Pre-I-1 & Precision of data values                & G        &           &        &        \\ \hline
Recoverability     &                                  & Rec-D-1                         & Data recoverability ratio               & G        &           &        &        \\ \hline
Relevance          & Rel-ML-1                         &                                 & Feature relevance                       & K        & x         & x      &        \\ \hline
Relevance          & Rel-ML-2                         &                                 & Record relevance                        & K        & x         & x      &        \\ \hline
Representativeness & Rep-ML-1                         &                                 & Representativeness ratio                & N        &           & x      &        \\ \hline
Similarity         & Sim-ML-1                         &                                 & Sample similarity                       & K        &           &        & x      \\ \hline
Similarity         & Sim-ML-2                         &                                 & Tightness                               & K        &           &        & x      \\ \hline
Similarity         & Sim-ML-3                         &                                 & Sample independency                     & K        &           &        & x      \\ \hline
\rowcolor{impl}Timeliness         & Tml-ML-1                         &                                 & Timeliness                              & K        & x         & x      &        \\ \hline
Understandability  & \cellcolor[HTML]{C0C0C0}Und-ML-1 & \cellcolor[HTML]{C0C0C0}Und-I-1 & Symbols understandability               & N        &           & x      &        \\ \hline
Understandability  & \cellcolor[HTML]{C0C0C0}Und-ML-2 & \cellcolor[HTML]{C0C0C0}Und-I-2 & Semantic understandability              & N        &           & x      &        \\ \hline
Understandability  & \cellcolor[HTML]{C0C0C0}Und-ML-3 & \cellcolor[HTML]{C0C0C0}Und-I-4 & Data value understandability            & N        &           & x      &        \\ \hline
Understandability  & \cellcolor[HTML]{C0C0C0}Und-ML-4 & \cellcolor[HTML]{C0C0C0}Und-D-2 & Data representation underst.            & N        &           & x      &        \\ \bottomrule
\end{tabular}%
}

\label{table:allDQMs}
\end{table*}

%% file: figures/guidelines-v2.tex
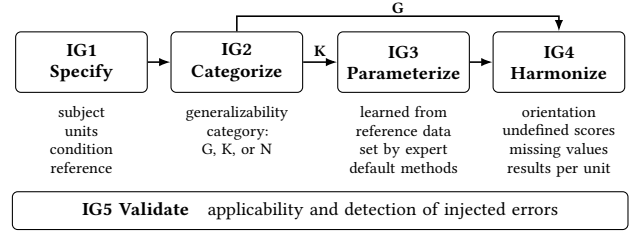
\begin{figure}[t]
  \centering
  \begin{tikzpicture}[
      node distance=3mm,
      ig/.style={draw, rounded corners=2pt, text width=1.6cm, minimum height=8mm, inner sep=2pt, align=center, font=\footnotesize\bfseries},
      igsub/.style={text width=1.6cm, inner sep=2pt, align=center, font=\scriptsize},
      igbar/.style={draw, rounded corners=2pt, text width=8.0cm, minimum height=5mm, inner sep=2pt, align=center, font=\footnotesize},
      arr/.style={-{Latex[length=1.5mm]}, semithick},
      lbl/.style={font=\scriptsize\bfseries, inner sep=1pt}
    ]
    \node[ig] (g1) {IG1\\Specify};
    \node[ig, right= of g1] (g2) {IG2\\Categorize};
    \node[ig, right=4.5mm of g2] (g3) {IG3\\Parameterize};
    \node[ig, right= of g3] (g4) {IG4\\Harmonize};

    \node[igsub, below=1mm of g1] {subject\\units\\condition\\reference};
    \node[igsub, below=1mm of g2] {generalizability category:\\G, K, or N};
    \node[igsub, below=1mm of g3] {learned from reference data\\set by expert\\default methods};
    \node[igsub, below=1mm of g4] {orientation\\undefined scores\\missing values\\results per unit};

    \draw[arr] (g1) -- (g2);
    \draw[arr] (g2) -- node[lbl, above, yshift=1pt] {K} (g3);
    \draw[arr] (g3) -- (g4);

    \coordinate (a) at ($(g2.north)+(0,2mm)$);
    \coordinate (b) at ($(g4.north)+(0,2mm)$);
    \draw[arr] (g2.north) -- (a) -- node[lbl, above] {G} (b) -- (g4.north);

    \path (g1.south west) -- (g4.south east) coordinate[midway] (mid);
    \node[igbar, below=13mm of mid] (g5) {\textbf{IG5 Validate}\quad applicability and detection of injected errors};
  \end{tikzpicture}
  \caption{Overview of the five implementation guidelines. IG1--IG4 turn a textual DQM definition into an executable DQM; DQMs of category G need no parameters and skip IG3. The result is validated in IG5.}
  \Description{Four boxes in a row, connected by arrows: Specify (subject, units, condition, reference), Categorize (category G, K or N), Parameterize (from reference data, by an expert, or by default) and Harmonize (orientation, undefined scores, missing values, results per unit). The arrow from Categorize to Parameterize is labelled K; a second arrow labelled G leads from Categorize over Parameterize directly to Harmonize. A fifth box below, Validate, covers applicability and detection of injected errors.}
  \label{fig:guidelines}
\end{figure}

%% file: sections/4_implementation.tex
\section{A Reference Implementation for DQMs}
\label{sec:implementation}
To illustrate the utility of the implementation guidelines, we applied them to the \numDQMs{} DQMs selected in \autoref{sec:dqm-classification} and provide the result as the open-source library \ourpackage. 
In the following, we describe how \ourpackage realizes IG1--IG4, while IG5 is covered in the evaluation in \autoref{sec:eval}. 
\ourpackage follows the scikit-learn API~\cite{sklearn_api} and operates on Narwhals-compatible dataframes~\cite{narwhals}.

\textbf{Specify.} 
For each metric, we specified subject, unit, condition and reference. 
The subject is selected when a DQM is created: a DQM on column level takes the name of the column, e.g., \texttt{DataAccuracyRange("temperature")}, whereas a DQM on table level, such as \texttt{RecordCompleteness()}, takes no column. 
For each DQM, \ourpackage documents the four components of Definition~\ref{def:dqm} as well as the remaining specifications that ISO/IEC 25021 requires, such as the scale type~\cite{iso25021}. The implementation thus covers all mandatory specifications of the standard.

\textbf{Categorize.} According to \autoref{table:allDQMs}, seven of the \numDQMs{} DQMs are of category G and can be used directly, while thirteen are of category K and need to be parameterized.

\textbf{Parameterize.}
The parameters of a category K DQM are obtained in one of two ways.
First, users can manually set parameters when initializing the DQM, allowing experts to use their domain knowledge.
Second, the \texttt{fit} method then learns all remaining parameters from a clean instance of the dataset, i.e., the reference data.
Learnable parameters range from simple derivations, such as the accepted domain values for Syntactic data accuracy (Acc-I-1), to more advanced methods, such as mining semantic rules in the form of Boolean expressions for Semantic consistency (Con-I-6). 
For Semantic data accuracy (Acc-I-2), \texttt{fit} samples exemplary records from the reference data, which an LLM uses to judge each value in the context of its record, following LLM-based error detection~\cite{narayanCanFoundationModels2022,Glock_2025}.
In addition, \texttt{method} and \texttt{threshold} are design choices rather than learnable quantities. 
For instance, for outlier detection, \texttt{method} selects the detection procedure, e.g., the rule based on the median absolute deviation (MAD), while \texttt{threshold} determines how far a value may deviate from the median, in multiples of the MAD, before it is flagged as an outlier. 
Each design choice has a default, such that a DQM can be used without any manual configuration. Consequently, no rules need to be defined manually to apply the ISO DQMs to a new dataset: a clean reference dataset is sufficient.

\textbf{Harmonize.}
The \texttt{score} method accepts the data instance to be measured and returns the score of a metric, which follows the conventions from \autoref{sec:model}: all scores are in $[0,1]$, a higher value indicates better DQ, and an undefined score is reported as \texttt{NaN}.
Accordingly, \ourpackage reports $1-X$ for DQMs that the standards define as a risk, e.g., Risk of data set inaccuracy (Acc-I-4).
For DQMs whose units are cells or rows, \texttt{predict} additionally returns the condition result per unit; for Value completeness (Com-ML-1), these per-row results are the scores of Record completeness (Com-I-1), so \ourpackage implements both in one DQM.

\textbf{DQMs beyond the model.} Distribution of data values (Con-ML-2) is not covered by Definition~\ref{def:dqm}, since its measurement function compares the distribution of an entire column against a reference distribution. Here, \texttt{fit} learns the reference distribution, and \texttt{score} reports $1$ minus the distance between the reference and the observed distribution, which is in $[0,1]$. Results per unit do not exist for this DQM.

%% file: sections/5_evaluation.tex
\section{Evaluation and Results}
\label{sec:eval}
\begin{figure}
    \includegraphics{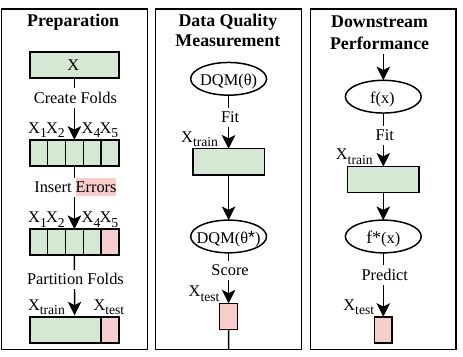}
    \caption{Experimental procedure: errors are injected into one fold of the dataset (\textit{left}); DQMs learn their parameters on the clean folds and score the corrupted fold (\textit{middle}), and an ML model is trained and evaluated on the same folds (\textit{right}).}
    \Description{Three panels. Left: a dataset is split into five folds, errors are inserted into one fold, and that fold becomes the test set while the other four form the training set. Middle: a DQM is fit on the training set and scores the test set. Right: an ML model is fit on the training set and predicts on the test set.}
    \label{fig:experiment-setup}
\end{figure}

In this section, we apply IG5 to the DQMs of \ourpackage: a DQM is validated if its score decreases monotonically with errors that belong to its own DQ dimension and remains stable for errors of other dimensions. We conduct three experiments. 
First, we test both criteria by injecting errors of individual types into a real-world dataset (\autoref{subsec:data-quality-measurement}).
Second, we show that DQM scores decline together with the performance of a downstream ML task (\autoref{subsec:downstream}).
Third, we show that the DQMs scale linearly with the size of the data, on a dataset with 41 million rows (\autoref{subsec:runtime}).
In all experiments, we follow the experimental procedure in \autoref{fig:experiment-setup}: we split a dataset into five folds, inject errors with increasing error rates into one fold, and use the remaining folds as clean training data (left). The DQMs learn their parameters on the training folds and score the corrupted test fold (middle); in the downstream experiment, an ML model is trained and evaluated on the same folds (right). We repeat this for each fold and report means and standard deviations over the five repetitions.

\subsection{Data Quality Measurement}
\label{subsec:data-quality-measurement}
The first experiment demonstrates that the DQMs defined in \ourpackage react as expected to data errors and that the DQM scores decrease monotonically with increasing amounts of errors injected into the data. 

\paragraph{Dataset}
To compare with prior work, we illustrate the capabilities of the DQMs defined in \ourpackage for measuring data quality on the HOSP dataset~\cite{xuchuHolisticDataCleaning2013}. While this dataset is not the most realistic one, it has become a well established benchmark for data cleaning~\cite{mahdaviBaranEffectiveError2020,jung2025realistic,rekatsinasHoloCleanHolisticData2017,narayanCanFoundationModels2022}.
It consists of 1,000 rows and 20 columns, all of which hold string values.
To prepare the experiment, we transform the columns \textit{Score} (40\% $\rightarrow 40$) and \textit{Sample} (26 patients $\rightarrow 26$) into numerical ones.
We also drop the columns \textit{index}, which holds an unused ordinal index, and \textit{Address2} and \textit{Address3}, which both contain the constant value ``empty''.

\paragraph{Cross-validation}
To obtain robust and realistic measurements, we apply cross-validation and  randomly partition the dataset into five equal folds containing $200$ rows each, and carry out five measurements in a leave-one-fold-out fashion for each configuration: four folds form the training data $X_{\text{train}}$, and the remaining fold is the test data $X_{\text{test}}$ (see~\autoref{fig:experiment-setup}).
The training data is never corrupted with errors.%

\paragraph{Error injection}
We inject errors into $X_{\text{test}}$ with \taberr by defining six error scenarios following~\cite{jung2025realistic}, which we summarize in~\autoref{tab:error-scenarios}.
We limit every error scenario to a single error type to isolate its effect on the DQMs. The last column of~\autoref{tab:error-scenarios} lists the DQ dimensions each scenario violates.
Following the error models of \citet{jung2025realistic}, the  scenarios distribute their errors completely at random (ECAR), i.e., independently of the data. The only exception is \texttt{blank record}: to empty entire rows rather than individual cells, its errors depend on the first column (error at random, EAR). Details on the error models are provided in~\cite{jung2025realistic}. 
\begin{table}[t]
  \centering
  \footnotesize
  \renewcommand{\arraystretch}{1.25}
  \caption{%
    Error scenarios injected into $X_{\text{test}}$, each limited to one error type with columns and DQ dimensions affected.
  }
  \label{tab:error-scenarios}
  \begin{threeparttable}
    \begin{tabular}{@{}lp{0.40\columnwidth}ll@{}}
      \toprule
      Scenario & Description & Cols. & Dimension\tnote{a} \\
      \midrule
      \texttt{blank record}
        & Empties entire records; error rate is the share of emptied records
        & 17 & Compl. \\
      \texttt{missing}
        & Removes individual cells
        & 17 & Compl. \\
      \texttt{swap}
        & Replaces a value with another from the same column
        & 14 str\tnote{b} & Acc., Cons. \\
      \texttt{typo}
        & Keyboard-adjacent misspellings
        & 15 str & Acc., Cons. \\
      \texttt{outlier}
        & Pushes values past the interquartile fence and adds Gaussian noise
        & 2 num & Acc. \\
      \texttt{wrong unit}
        & Rescales values by a factor of $1/100$
        & 2 num & Acc. \\
      \bottomrule
    \end{tabular}
    \begin{tablenotes}
      \footnotesize
      \item[a] accuracy, completeness, consistency
      \item[b] \textit{HospitalType} excluded because it is constant.
    \end{tablenotes}
  \end{threeparttable}
\end{table}
We control the amount of noise in the dataset by setting the per-column error rate to 1\%, 5\%, 10\%, 25\%, and 50\% for each error scenario, generating increasingly noisy test datasets $X_{\text{test}}$.
We additionally carry out a \texttt{clean} run where we do not inject errors, which serves as the baseline every error scenario is compared against.
\begin{table*}
  \centering
  \setlength{\tabcolsep}{4pt}
  \caption{
    Average change in DQM scores from clean data to a 50\% per-column error rate.
    Since each error scenario corrupts multiple columns, we report the average of the change in DQM score over the affected columns, with the standard deviation over five folds.
    Changes of at least $0.05$ are bold, and a dash marks that a DQM could not be evaluated under this scenario.
  }
    \label{tab:overview-data-quality}
  \begin{tabular}{lrrrrrrr}
    \toprule
    & & \multicolumn{6}{c}{Error scenario} \\
    \cmidrule(lr){3-8}
    DQ metric & clean & \texttt{missing} & \texttt{blank record} & \texttt{typo} & \texttt{swap} & \texttt{outlier} & \texttt{wrong unit} \\
    \midrule
    Data accuracy range & 0.999 & +0.001{\scriptsize$\pm0.001$} & +0.001{\scriptsize$\pm0.001$} & -- & -- & \textbf{-0.348}{\scriptsize$\pm0.013$} & +0.000{\scriptsize$\pm0.000$} \\
    Risk of data set inaccuracy & 0.903 & +0.004{\scriptsize$\pm0.017$} & +0.011{\scriptsize$\pm0.024$} & -- & -- & \textbf{-0.404}{\scriptsize$\pm0.042$} & \textbf{-0.204}{\scriptsize$\pm0.009$} \\
    Semantic data accuracy & 0.983 & -0.006{\scriptsize$\pm0.006$} & -0.005{\scriptsize$\pm0.005$} & \textbf{-0.475}{\scriptsize$\pm0.013$} & \textbf{-0.567}{\scriptsize$\pm0.031$} & \textbf{-0.517}{\scriptsize$\pm0.019$} & \textbf{-0.465}{\scriptsize$\pm0.023$} \\
    Syntactic data accuracy & 0.986 & +0.001{\scriptsize$\pm0.003$} & -0.001{\scriptsize$\pm0.001$} & \textbf{-0.488}{\scriptsize$\pm0.002$} & -0.004{\scriptsize$\pm0.001$} & \textbf{-0.448}{\scriptsize$\pm0.013$} & \textbf{-0.416}{\scriptsize$\pm0.012$} \\
    \addlinespace
    Empty records & 1.000 & +0.000{\scriptsize$\pm0.000$} & \textbf{-0.500}{\scriptsize$\pm0.000$} & +0.000{\scriptsize$\pm0.000$} & +0.000{\scriptsize$\pm0.000$} & +0.000{\scriptsize$\pm0.000$} & +0.000{\scriptsize$\pm0.000$} \\
    Feature completeness & 0.987 & \textbf{-0.493}{\scriptsize$\pm0.001$} & \textbf{-0.493}{\scriptsize$\pm0.001$} & +0.000{\scriptsize$\pm0.000$} & +0.000{\scriptsize$\pm0.000$} & +0.000{\scriptsize$\pm0.000$} & +0.000{\scriptsize$\pm0.000$} \\
    Label completeness & 1.000 & \textbf{-0.500}{\scriptsize$\pm0.000$} & \textbf{-0.500}{\scriptsize$\pm0.000$} & +0.000{\scriptsize$\pm0.000$} & +0.000{\scriptsize$\pm0.000$} & -- & -- \\
    Record completeness & 0.833 & \textbf{-0.833}{\scriptsize$\pm0.047$} & \textbf{-0.416}{\scriptsize$\pm0.020$} & +0.000{\scriptsize$\pm0.000$} & +0.000{\scriptsize$\pm0.000$} & +0.000{\scriptsize$\pm0.000$} & +0.000{\scriptsize$\pm0.000$} \\
    Value completeness & 0.987 & \textbf{-0.493}{\scriptsize$\pm0.001$} & \textbf{-0.493}{\scriptsize$\pm0.001$} & +0.000{\scriptsize$\pm0.000$} & +0.000{\scriptsize$\pm0.000$} & +0.000{\scriptsize$\pm0.000$} & +0.000{\scriptsize$\pm0.000$} \\
    Value occurrence completeness & 0.839 & \textbf{-0.366}{\scriptsize$\pm0.005$} & \textbf{-0.397}{\scriptsize$\pm0.007$} & \textbf{-0.373}{\scriptsize$\pm0.006$} & \textbf{-0.142}{\scriptsize$\pm0.016$} & \textbf{-0.255}{\scriptsize$\pm0.021$} & \textbf{-0.235}{\scriptsize$\pm0.022$} \\
    \addlinespace
    Data format consistency & 1.000 & -0.000{\scriptsize$\pm0.000$} & +0.000{\scriptsize$\pm0.000$} & \textbf{-0.169}{\scriptsize$\pm0.005$} & -0.001{\scriptsize$\pm0.001$} & -- & -- \\
    Data record consistency & 1.000 & -- & +0.000{\scriptsize$\pm0.000$} & +0.000{\scriptsize$\pm0.000$} & +0.000{\scriptsize$\pm0.000$} & +0.000{\scriptsize$\pm0.000$} & +0.000{\scriptsize$\pm0.000$} \\
    Data value distribution & 0.876 & -0.043{\scriptsize$\pm0.008$} & \textbf{-0.152}{\scriptsize$\pm0.018$} & \textbf{-0.373}{\scriptsize$\pm0.006$} & \textbf{-0.142}{\scriptsize$\pm0.016$} & \textbf{-0.283}{\scriptsize$\pm0.019$} & \textbf{-0.381}{\scriptsize$\pm0.032$} \\
    Risk of data inconsistency & 0.037 & +0.045{\scriptsize$\pm0.005$} & +0.021{\scriptsize$\pm0.006$} & \textbf{+0.376}{\scriptsize$\pm0.006$} & +0.003{\scriptsize$\pm0.003$} & \textbf{+0.419}{\scriptsize$\pm0.023$} & \textbf{+0.099}{\scriptsize$\pm0.018$} \\
    Semantic consistency & 1.000 & +0.000{\scriptsize$\pm0.000$} & +0.000{\scriptsize$\pm0.000$} & \textbf{-0.509}{\scriptsize$\pm0.008$} & \textbf{-0.887}{\scriptsize$\pm0.005$} & -- & -- \\
    \bottomrule
  \end{tabular}
\end{table*}
Every DQM that requires parameters estimates them on the clean $X_{\text{train}}$ and every DQM then scores $X_{\text{test}}$, as depicted in the middle panel in~\autoref{fig:experiment-setup}. 
The DQ measurement results are shown in~\autoref{tab:overview-data-quality}. %
The first column shows the DQ score on the error-free test data, averaged over the five folds and all columns of the table the DQM applies to.
Each subsequent column corresponds to one error scenario.
It shows the change in DQ score from the error-free to the corrupted test data at an error rate of $50\%$, averaged over the five folds on the columns the scenario corrupts.\footnote{The first column and the reported change are not additive, because the change is measured against the error-free score on the columns the error scenario corrupts. When a scenario corrupts only some of the columns a DQM is applied to, this score differs from the error-free score reported in the first column.}
We report the change in score rather than the score itself, because a DQM does not necessarily score $1.00$ on clean data: Value occurrence completeness, for instance, scores $0.839$ on the clean HOSP table, since value frequencies differ between folds even without errors. Moreover, HOSP is a real-world dataset and may contain errors of its own; the training data is clean only in the sense that we did not inject errors.

\paragraph{Impact of errors on DQMs}
The DQMs can be grouped by the number of error scenarios they react to, that is, by the number of bold entries in their row of \autoref{tab:overview-data-quality}.

\textsc{Four to six scenarios.}
Value occurrence completeness responds to all six error scenarios, Data value distribution to five, and the LLM-based Semantic data accuracy to four.
This is the case because Value occurrence completeness and Data value distribution both derive their scores by learning a column distribution in $X_{\text{train}}$ and then comparing it to the one observed in $X_{\text{test}}$:
most error scenarios measurably affect the column distribution.
For the same reason, \texttt{swap} affects both DQMs only moderately ($-0.142$ each): replacing a value with another one from the same column often has negligible impact on the distribution.
Semantic data accuracy, on the other hand, reacts broadly because it checks whether a value equals the true value it should represent.
The LLM powering the DQM, DeepSeek V4 Flash 0731~\cite{deepseek-aiDeepSeekV4HighlyEfficient2026}, detects the semantically incorrect values of every error scenario that produces non-empty values.

\textsc{Three scenarios.}
Syntactic data accuracy and Risk of data inconsistency react to three error scenarios each.
Syntactic data accuracy learns the set of distinct values observed in $X_{\text{train}}$ and by design cannot detect \texttt{swap} errors, because the swapped values are already part of the learned set.
Risk of data inconsistency measures the ratio of cells whose value occurs exactly once in their column, so the fewer unique values a column has, the worse the score becomes.
The score of $0.037$ on the clean data reveals that HOSP contains almost no unique values per column.
Because \texttt{typo}, \texttt{outlier}, and \texttt{wrong unit} errors alter affected values in place, they create new unique values, thereby causing the score to rise --- it is the only DQM whose score rises as the data degrades.

\textsc{Two scenarios.}
Except for Value occurrence completeness, the completeness DQMs react only to missing values, i.e.,~to \texttt{missing} and \texttt{blank record}.
Risk of data set inaccuracy detects outliers by inferring the median and the scaled median absolute deviation of $X_{\text{train}}$; it registers \texttt{outlier} ($-0.404$) and \texttt{wrong unit} ($-0.204$).
Semantic consistency mines the unary functional dependencies that hold between columns of $X_{\text{train}}$ and evaluates them per row on $X_{\text{test}}$.
It reacts to \texttt{typo} ($-0.509$) and \texttt{swap} ($-0.887$).

\textsc{One scenario.}
The DQMs Data accuracy range and Data format consistency react to one type of error each in our measurement.
Data accuracy range is only defined for numerical columns and verifies that values lie within a valid interval.
It reacts only to \texttt{outlier} ($-0.348$): \texttt{wrong unit} rescales affected values by a factor of $1/100$, which keeps them within the valid interval observed on $X_{\text{train}}$.
Data format consistency, in turn, is defined only for string columns, and reacts solely to \texttt{typo} ($-0.169$), because no other error scenario violates the format of a column.

\textsc{No scenario.}
Finally, Data record consistency reacts to none of the error scenarios.
It measures the share of duplicate records, and none of our scenarios creates duplicates.
Under \texttt{missing} at a $50\%$ error rate, every row contains a missing value, meaning that the DQM cannot be evaluated.

\begin{figure*}
    \includegraphics{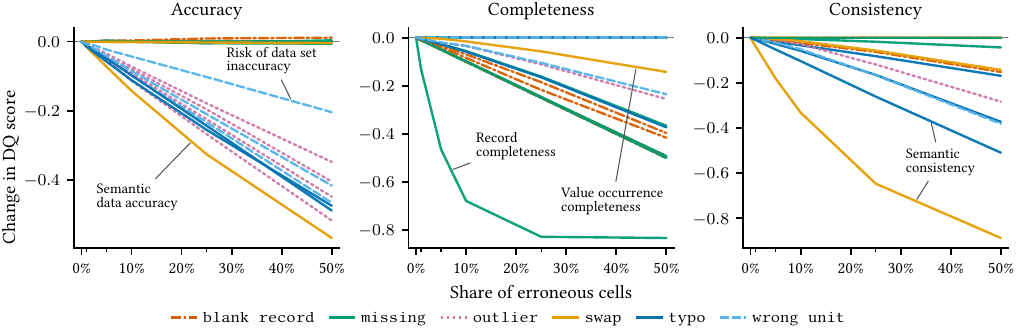}
    \caption{%
    The DQ scores drop or stay constant as the error rate increases on the HOSP dataset.
    Each line represents one DQM's reaction to the error scenario listed in the legend.
    Measurements are grouped by data quality dimension.
    Risk of data inconsistency is excluded from the graph for conciseness.
   }
   \Description{Three line plots, for accuracy, completeness and consistency, show the change in DQ score as the share of erroneous cells rises from 0 to 50 percent. Most lines fall roughly linearly and a few stay at zero. Swapped values impact Semantic data accuracy and Semantic consistency most, down by about 0.55 and 0.9. Record completeness falls to about -0.83 by 25 percent missing values and then levels off.}
    \label{fig:data-quality-measurement}
\end{figure*}

\medskip
\autoref{fig:data-quality-measurement} shows, per DQ dimension, that the scores either fall monotonically as the error rate grows from $0$ to $50\%$, or stay constant.
We omit Risk of data inconsistency, whose rising change in DQ score would distort the scale of \autoref{fig:data-quality-measurement}: it scores the share of unique values per column, and the injected errors replace values by new, unique ones, so the score rises. This is not a flaw of the implementation, but of the metric's assumption, which holds for identifiers, not for the categorical columns of HOSP.

In summary, DQMs in \ourpackage track deteriorating DQ without manual input:
eight of the fifteen DQMs learn their parameters from clean reference data, seven need none, and every error scenario is detected by at least four DQMs.
However, expert knowledge is still needed to decide which DQMs suit a dataset: Risk of data inconsistency, for instance, measures how unique a column's values are, which says little about DQ on HOSP. 

\subsection{Data Quality Impact on ML Tasks}
\label{subsec:downstream}
The second experiment tests whether the DQM scores decline together with the performance of a downstream ML task, i.e., whether the DQMs detect data errors that harm an application. 

\paragraph{Dataset} For the downstream impact experiments we use SynTabFall, a synthetic health dataset released for training fall risk assessment ML models~\cite{nanevskiAnonymizedUsefulSynthetic2026}.
It has 745,380 rows and 44 columns, where each row is a patient record including demographics, diagnoses, and medical care status.
The task is to predict whether a patient fell, a binary classification of the column \texttt{fallen}.

\paragraph{Cross-validation}
We follow the same procedure as in \autoref{subsec:data-quality-measurement} and split the dataset into five folds of about 149,000 rows each.

\paragraph{Error injection}
We inject five different realistic error types, which together form one error scenario with a common error rate:
\begin{enumerate}

\item Incomplete join: a job that left-joins medical attributes onto patient records and fails midway, leaving \texttt{procedure} (OPS  codes) and \texttt{medical\_items} columns (e.g.,~ventilation mask use) empty for a share of rows (ECAR).

\item Incomplete documentation: nursing assessments are unrecorded in certain age groups (EAR on \texttt{age}), affecting the  eleven nursing care, seven cognition, and four walking attributes as defined in \citet{nanevskiAnonymizedUsefulSynthetic2026}.

\item Incorrect cells: in \texttt{fall-risk}, whose three levels encode no risk \texttt{0}, increased risk \texttt{+} and high risk \texttt{++}, we swap each affected cell to an incorrect category (ECAR), simulating a miscoded patient intake form.

\item Age shift: 100 years are subtracted from \texttt{age} for a subset of rows (error not at random, ENAR), modeling a system that infers the wrong century when only two digits of the birth year are stored; for example, 1926 becomes 2026.

\item Typos: characters in ICD-10-GM codes of \texttt{diagnosis} are corrupted (ENAR), e.g., turning \texttt{S022} into \texttt{S032} or \texttt{R51} into \texttt{Rt1}.
This simulates one corrupted region of the diagnosis code catalog.

\end{enumerate}

We do not evaluate the Label completeness DQM, since the target \texttt{fallen} stays intact and would always score $1.0$.
We also omit Semantic data accuracy: it issues one LLM request per cell, so its cost grows linearly with the table, amounting to roughly $3.3 \times 10^7$ for a single pass over the table.

\paragraph{ML model training}
We fit scikit-learn's~\cite{scikit-learn} \emph{HistGradientBoostingClassifier}, a gradient-boosted decision tree ensemble, with 200 boosting iterations on $X_{\text{train}}$, and predict the binary target column \texttt{fallen}.
We ordinal-encode the categorical columns, treat missing values as a category of their own, and let the model natively handle missing values in the single numeric column \texttt{age}.

The classification task is strongly imbalanced: the positive class, patients who fell, makes up only $0.98\%$ of rows.
We tune the classifier's decision threshold for macro-$F_1$ score by five-fold cross-validation on $X_{\text{train}}$ to reproduce the reference setup.
On the error-free test data, the classifier reaches a macro-$F_1$ score of $0.577${\scriptsize$\pm0.002$}, comparable to the $0.56$ that \citet{nanevskiAnonymizedUsefulSynthetic2026} report for a model trained on the same synthetic data but tested on real patient records.
To measure how errors affect the classifier, we report the area under the ROC curve (ROC-AUC), which evaluates how well the classifier ranks patients independent of the decision threshold and the class ratio.

\begin{figure}
    \includegraphics{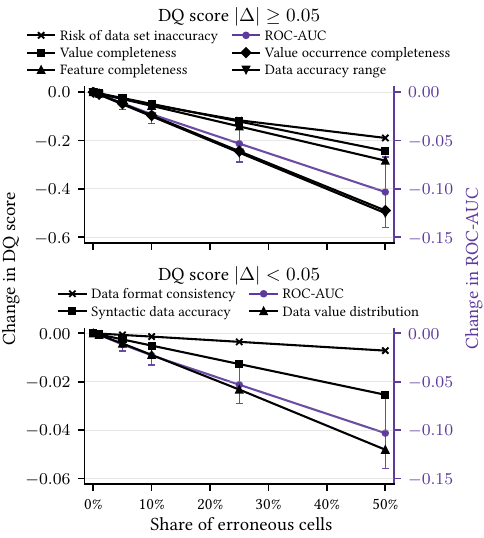}
    \caption{%
    On the fall risk prediction ML task (SynTabFall), the scores of the DQMs that react to the errors fall together with the classifier's ROC-AUC as the error rate rises.
    The left panel shows DQMs whose score changes by $0.05$ or more, the right panel those that change less.
    Whiskers show the standard deviation of ROC-AUC across folds.
   }
   \Description{Two stacked line plots show the change in DQ score (left axis) and in ROC-AUC (right axis) as the share of erroneous cells rises from 0 to 50 percent. In the top plot, five DQMs fall by between 0.2 and 0.5. In the bottom plot, three DQMs fall by less than 0.05. In both, ROC-AUC falls in step to about -0.10.}
    \label{fig:downstream-impact}
\end{figure}

\paragraph{Impact of errors on DQMs and ML task}
\autoref{tab:downstream-performance} lists the classifier's ROC-AUC alongside the DQM scores for increasing error rates.
The ROC-AUC falls monotonically from $0.879${\scriptsize$\pm0.003$} on the clean test data to $0.776${\scriptsize$\pm0.036$} when half the cells are corrupted, showing that the error scenario has a downstream cost.
Five DQMs react strongly to the scenario (left panel of \autoref{fig:downstream-impact}):
Data accuracy range's score drops by $0.500$ because ages shifted back by 100 years fall outside the valid range, and Risk of data set inaccuracy ($-0.189$) detects the shifted ages as outliers.
The completeness DQMs Value occurrence completeness ($-0.489$), Feature completeness ($-0.283$) and Value completeness ($-0.243$) fall in proportion to the injected missing values.

\begin{table*}[t]
  \centering
  \setlength{\tabcolsep}{4pt}
  \caption{%
    Mean DQ score over the columns the error scenario damages on the SynTabFall dataset, averaged over five folds ($\pm$ standard deviation).
    The first row is the classifier's ROC-AUC.
    $\Delta$ is the change from $0\%$ to $50\%$ (bold at $0.05$ or more), and a dash marks a DQM that returned no score.
  }
    \label{tab:downstream-performance}
  \begin{tabular}{lrrrrrrr}
    \toprule
    & \multicolumn{6}{c}{Share of erroneous cells} & \\
    \cmidrule(lr){2-7}
    Data quality measure & 0\% & 1\% & 5\% & 10\% & 25\% & 50\% & $\Delta$ \\
    \midrule
    ROC-AUC & $0.879${\scriptsize$\pm0.003$} & $0.878${\scriptsize$\pm0.004$} & $0.868${\scriptsize$\pm0.007$} & $0.856${\scriptsize$\pm0.010$} & $0.826${\scriptsize$\pm0.019$} & $0.776${\scriptsize$\pm0.036$} & $\mathbf{-0.103}$ \\
    \midrule
    Data accuracy range & $1.000${\scriptsize$\pm0.000$} & $0.990${\scriptsize$\pm0.000$} & $0.950${\scriptsize$\pm0.000$} & $0.900${\scriptsize$\pm0.000$} & $0.750${\scriptsize$\pm0.000$} & $0.500${\scriptsize$\pm0.000$} & $\mathbf{-0.500}$ \\
    Risk of data set inaccuracy & $1.000${\scriptsize$\pm0.000$} & $0.994${\scriptsize$\pm0.005$} & $0.972${\scriptsize$\pm0.026$} & $0.948${\scriptsize$\pm0.050$} & $0.884${\scriptsize$\pm0.113$} & $0.811${\scriptsize$\pm0.139$} & $\mathbf{-0.189}$ \\
    Syntactic data accuracy & $1.000${\scriptsize$\pm0.000$} & $1.000${\scriptsize$\pm0.000$} & $0.997${\scriptsize$\pm0.000$} & $0.995${\scriptsize$\pm0.000$} & $0.987${\scriptsize$\pm0.001$} & $0.975${\scriptsize$\pm0.001$} & $-0.025$ \\
    \addlinespace
    Empty records & $1.000${\scriptsize$\pm0.000$} & $1.000${\scriptsize$\pm0.000$} & $1.000${\scriptsize$\pm0.000$} & $1.000${\scriptsize$\pm0.000$} & $1.000${\scriptsize$\pm0.000$} & $1.000${\scriptsize$\pm0.000$} & $0.000$ \\
    Feature completeness & $0.628${\scriptsize$\pm0.000$} & $0.622${\scriptsize$\pm0.000$} & $0.599${\scriptsize$\pm0.000$} & $0.571${\scriptsize$\pm0.000$} & $0.486${\scriptsize$\pm0.001$} & $0.345${\scriptsize$\pm0.001$} & $\mathbf{-0.283}$ \\
    Record completeness & $0.000${\scriptsize$\pm0.000$} & $0.000${\scriptsize$\pm0.000$} & $0.000${\scriptsize$\pm0.000$} & $0.000${\scriptsize$\pm0.000$} & $0.000${\scriptsize$\pm0.000$} & $0.000${\scriptsize$\pm0.000$} & $0.000$ \\
    Value completeness & $0.583${\scriptsize$\pm0.000$} & $0.578${\scriptsize$\pm0.000$} & $0.559${\scriptsize$\pm0.000$} & $0.535${\scriptsize$\pm0.000$} & $0.462${\scriptsize$\pm0.001$} & $0.340${\scriptsize$\pm0.001$} & $\mathbf{-0.243}$ \\
    Value occurrence completeness & $0.996${\scriptsize$\pm0.001$} & $0.987${\scriptsize$\pm0.001$} & $0.949${\scriptsize$\pm0.002$} & $0.900${\scriptsize$\pm0.003$} & $0.752${\scriptsize$\pm0.006$} & $0.507${\scriptsize$\pm0.008$} & $\mathbf{-0.489}$ \\
    \addlinespace
    Data format consistency & $1.000${\scriptsize$\pm0.000$} & $1.000${\scriptsize$\pm0.000$} & $0.999${\scriptsize$\pm0.000$} & $0.999${\scriptsize$\pm0.000$} & $0.996${\scriptsize$\pm0.000$} & $0.993${\scriptsize$\pm0.001$} & $-0.007$ \\
    Data record consistency & -- & -- & -- & -- & -- & -- & -- \\
    Data value distribution & $0.997${\scriptsize$\pm0.000$} & $0.996${\scriptsize$\pm0.000$} & $0.992${\scriptsize$\pm0.001$} & $0.988${\scriptsize$\pm0.001$} & $0.973${\scriptsize$\pm0.003$} & $0.949${\scriptsize$\pm0.007$} & $-0.048$ \\
    Risk of data inconsistency & $0.000${\scriptsize$\pm0.000$} & $0.000${\scriptsize$\pm0.000$} & $0.000${\scriptsize$\pm0.000$} & $0.000${\scriptsize$\pm0.000$} & $0.001${\scriptsize$\pm0.000$} & $0.001${\scriptsize$\pm0.000$} & $+0.001$ \\
    Semantic consistency & -- & -- & -- & -- & -- & -- & -- \\
    \bottomrule
  \end{tabular}
\end{table*}
Syntactic data accuracy ($-0.025$), Data format consistency ($-0.007$) and Data value distribution ($-0.048$) also react to the error scenario, but at a smaller magnitude, as the right panel of~\autoref{fig:downstream-impact} shows.
Each of the three judges one column at a time and therefore  registers only the errors in that column.
Syntactic data accuracy and Data format consistency detect the typos in \texttt{diagnosis}, and Syntactic data accuracy also flags the shifted \texttt{age}.
Data value distribution picks up the changed value frequencies in \texttt{diagnosis}, \texttt{fall-risk} and \texttt{age}.
\autoref{tab:downstream-performance} averages each DQM over all 36 columns the scenario damages.
In most of them, the scenario only introduces missing values, to which the three DQMs do not react, so averaging over all columns shrinks their changes.
These changes, however, are not noise: all three DQMs fall monotonically with the error rate, and at $50\%$ their drop is six to twenty-six times the standard deviation across folds.
Data record consistency only scores rows without missing values, and because every row in SynTabFall contains at least one missing value, it returns no score.
Record completeness reports $0$ for the same reason.
Semantic consistency finds no exact FD in the training data, meaning it returns no score either.
Empty records stays at $1.0$, since no row is emptied entirely, and Risk of data inconsistency stays at $0$, since the categorical columns of SynTabFall contain no unique values.

In summary, the experiment shows that \ourpackage's DQMs track the data damage that lowers the classifier's performance.
As expected for realistic data distributions, not all injected errors are reflected equally in the DQMs, nor in the  predictive performance of the ML model. 
Nevertheless, the DQMs that react to the errors, as well as the classifier's ROC-AUC, decrease monotonically with the error rate.

\subsection{Runtime Evaluation}
\label{subsec:runtime}
The third experiment tests whether the DQMs scale to large datasets, which is a prerequisite for continuous DQ monitoring in production settings~\cite{schelter_2018, Ehrlinger_2017}.

\paragraph{Dataset}
We evaluate the runtime characteristics of the DQMs defined in \ourpackage on the \textit{TLC Trip Record Data} of the New York City Taxi \& Limousine Commission from 2024~\cite{nycTLCTripRecordData}.
The dataset comprises 41,169,720 records, one per taxi trip, and 19 columns describing trip properties like the pickup and drop-off location, vendor, and applied rate.
It has been widely used for DBMS benchmarking in prior work~\cite{dingTsunamiLearnedMultidimensional2020,liuHowGoodAre2025}.
We concatenate the twelve monthly trip datasets for 2024 and shuffle the rows with a fixed seed to remove the ordering by pickup time, so that every sample is drawn from the same distribution.
Then, we join a zone lookup table on the pickup and drop-off location, hence adding six columns with exact FDs, for a total of 25 columns.

\paragraph{Experimental setup}
Before each measurement, we run all DQMs once untimed on $10^3$ rows, such that Python imports and first-call loading times are excluded.
The measurement runs on a machine with an AMD EPYC 7742 CPU with $2 \times 64$ cores, 256 hardware threads and 1TB\,RAM in a Kubernetes cluster.
Each measurement runs in an isolated environment limited to 4 CPU cores and 32\,GB of memory, or 96\,GB for the largest measurements, with Polars restricted to 4 threads.
Up to twelve containers run in parallel on the node.
We use Python 3.12, Polars 1.41.2, and Narwhals 2.22.1.
The runtimes we report use a Polars backend, and although \ourpackage also supports Pandas, we do not compare the two backends as prior work already established runtime~\cite{mozzilloEvaluationDataframeLibraries2025a} and sustainability~\cite{nahrstedtEmpiricalStudyEnergy2024} benchmarks.
The Label completeness DQM is excluded from the measurement, because the dataset has no ML target; the DQM shares its implementation with Feature completeness and scales identically.
We apply every table-level DQM to the table and every column-level DQM to each column it accepts, and time \texttt{fit} and \texttt{score} on $10^3, 3\times10^3, \dots, 3\times10^7$ and, finally, all 41,169,720 rows, with ten repetitions each.
The column-level DQMs are applied to multiple columns per run, and we report the median runtime.
We then take the average and calculate the standard deviation over the ten measurements.

\begin{figure*}
    \includegraphics{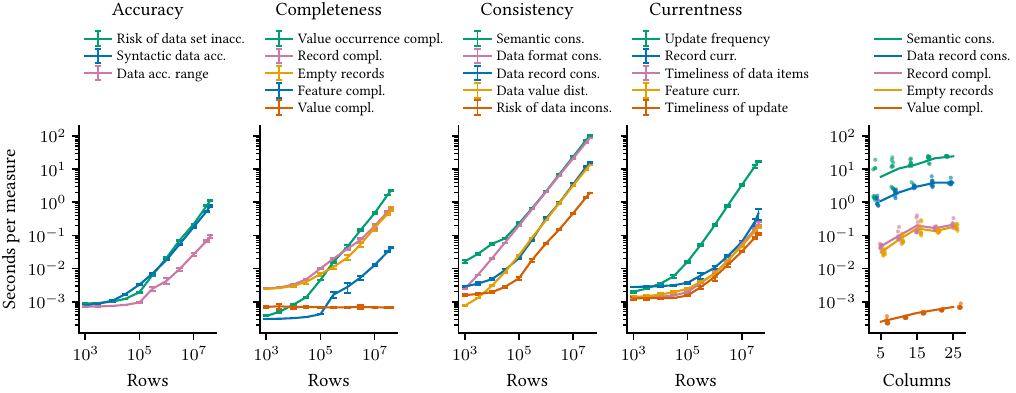}
    \caption{%
    DQMs' \texttt{fit} and \texttt{score} scale linearly with the number of rows and columns.
    The four panels on the left show that, below around $10^5$ rows, a fixed overhead per call dominates the runtime.
    The rightmost panel fixes $10^7$ rows and shows only the DQMs that read more than one column and are affected by the number of columns.
    Lines show the mean over ten repetitions, error bars the standard deviation.
   }
   \Description{Four log-log plots, one per DQ dimension accuracy, completeness, consistency and currentness, show seconds per measure against 1,000 to about 50 million rows. Most runtimes stay near one millisecond up to about 100,000 rows and then rise in linearly with the row count; the slowest DQMs, Semantic consistency and Data format consistency, reach about 100 seconds. A fifth plot fixes 10 million rows and shows runtime rising gradually from 5 to 25 columns for five DQMs.}
    \label{fig:scalability}
\end{figure*}

\paragraph{Data quality scoring scales linearly}
The four leftmost panels in~\autoref{fig:scalability} show that DQMs scale linearly with the number of rows.
Fifteen DQMs run as Polars queries and take at most $2.30 \pm 0.08$\,s on all rows (Value occurrence completeness), except for Data record consistency ($15.6 \pm 0.7$\,s), which hashes all columns of every row, and Update frequency ($17.2 \pm 0.9$\,s), which sorts columns and therefore grows slightly faster than linearly.
Considerably slower are three DQMs that run Python code in addition to Polars queries: Data format consistency ($86.5 \pm 0.4$\,s) compares each value to its format shape, Semantic consistency ($101 \pm 2$\,s) mines FDs in Python and checks them row by row, and Data value distribution computes the Kolmogorov--Smirnov statistic or the total variation distance, depending on the column's datatype.
Below around $10^5$ rows, the DQMs that only evaluate on Polars exhibit constant runtimes, because Narwhals translates the DQM into a Polars query, which Polars then optimizes and plans.
The three DQMs that require calculation in Python grow from $10^3$ to $10^4$ rows on, because their cost per value outweighs Polars' constant overhead early.
Value completeness runs in constant time, because Polars stores the number of missing values per column as metadata, and avoids scanning the rows for this reason.
The rightmost panel of~\autoref{fig:scalability} fixes $10^7$ rows, the smallest size from which the runtimes in the row panels grow linearly with the rows, and evaluates the table-scope DQMs and Semantic consistency on $5, 10, 15, 20, 25$ columns drawn in a new, random order in each repetition.
We exclude the Record currentness DQM from the evaluation, because it reads only columns containing timestamps, of which the dataset contains two.
The three table-scope DQMs Empty records, Record completeness, and Data record consistency evaluate one expression per column and combine the results per row, which causes their runtime to grow linearly with the number of columns.
Because Value completeness sums the missing value counts that Polars stores as metadata, its runtime grows with the columns, but not with the rows.
The dots in the panel show each measurement.
The spread of the table-scope DQMs does not shrink as the number of columns increases, although at 25 columns every repetition reads the same columns --- the spread is therefore run-to-run noise.

The spread of Semantic consistency, in contrast, depends on which columns are drawn.
Although it is a column-scope DQM, its \texttt{fit} checks every other column of the table for an FD, and its runtime grows linearly with the number of columns.
To mine FDs, the DQM collects distinct value pairs of two columns in a Python hash table, so its runtime also grows linearly with the number of distinct pairs.
Almost every value in the two timestamp columns is distinct, so nearly every row forms a distinct pair, causing the FD check against a timestamp column to dominate the runtime whenever one is drawn.
The more columns are drawn, the more likely both timestamp columns are included, until at 25 columns all of them do and the spread narrows to run-to-run noise.

\begin{figure}
    \includegraphics{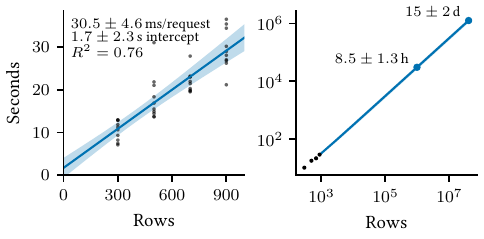}
    \caption{%
    Left: Semantic data accuracy's runtime over ten repetitions with the per-repetition linear fit and its 95\% confidence band.
    Right: the fit extrapolated to the full dataset.
   }
   \Description{Left: runtime of Semantic data accuracy at 300 to 900 rows, with ten repetitions per size. A linear fit gives 30.5 ± 4.6 ms per request with an intercept of 1.7 ± 2.3 seconds, R2 = 0.76. Right: the fit extended on log-log axes gives 8.5 ± 1.3 hours at about one million rows and 15 ± 2 days at about 40 million rows.}
    \label{fig:scalability-llm}
\end{figure}

The LLM-based Semantic data accuracy DQM sends one LLM request per distinct record, making it the most expensive DQM to evaluate, both in time and money.
We therefore exclude it from~\autoref{fig:scalability} and time \texttt{score} separately on the first $300, 500, 700$ and $900$ rows, with ten repetitions each.
The DQM queries the same LLM as in~\autoref{subsec:data-quality-measurement}, DeepSeek V4 Flash, with 32 concurrent requests.
All measured rows are distinct, and each repetition sends as many requests as there are rows.
Per repetition, we fit a line through the four row counts, where the slope is the time each additional request adds, and the intercept is the runtime independent of the number of requests (left panel in~\autoref{fig:scalability-llm}).
The fit yields $30.5 \pm 4.6$\,ms per request (95\% confidence intervals over the ten repetitions), which we extrapolate to $(1.25 \pm 0.19) \times 10^6$\,s for the full table (right panel).

In summary, all DQMs except Semantic data accuracy scale linearly and score the full table of 41 million rows in under two minutes. The LLM-based metric, at about 30\,ms per request, is limited to samples or small tables.

%% file: sections/6_discussion.tex
\section{Discussion and Open Challenges}
\label{sec:discussion}
In the following, we discuss the lessons learned when applying the guidelines, what enabled the automation, and the remaining limitations, before we outline two open challenges.

\textbf{Lessons from applying the guidelines.}
Applying the guidelines to the ISO metrics revealed where the standards require interpretation: several metrics delegate the choice of a method to the implementer, the term data item is not used consistently, and the orientation of the scores differs. We resolved these points (see IG1, IG3, and IG4), which we consider a prerequisite for comparing results across implementations.

Our evaluation further shows that an executable DQM is not necessarily a meaningful one for a given dataset. On HOSP, the score of Risk of data inconsistency increases with the number of errors, since the DQM counts unique values and errors such as typos create new ones. On SynTabFall, Record completeness is $0$ for all error rates, Data record consistency cannot be evaluated because every row contains a missing value, and Semantic consistency does not find a functional dependency in the reference data. These observations underline the importance of IG5: it needs to be stated for which data a DQM is applicable.
While many manual efforts around DQMs can be alleviated with a reference dataset, human expert knowledge remains a prerequisite for some, as classified in~\autoref{table:allDQMs}.

\textbf{Reference data as an enabler.}
To increase the degree of automation and allow for simple and automated creation of DQ tests compliant with ISO standards, we adopted a well-established principle in the ML community: rather than relying on manually defined rules for DQMs, we assume a clean reference set and learn the DQM parameters from that reference set. In addition to parameters that can be learned from the reference set, we also allow for more complex parameters for monitoring, e.g., semantic correctness with LLMs. 
A key ingredient of this automation is the reference data. Assuming a clean reference dataset is a substantial improvement over manually defined DQMs --- manual test definitions are often the bottleneck, and in many application scenarios, the lack of time needed to define and maintain rules is an important factor for why DQMs are neglected. The assumption of clean reference data is widely accepted, for instance in the ML community. In cases where no reference data is available or when there is reason to believe that the data is corrupted, we suggest falling back onto manually defined expert knowledge, and we support this option in our reference implementation. 

\textbf{Limitations.}
Our approach has two main limitations. First, a learned reference inherits the blind spots of the reference data: Syntactic data accuracy, for instance, cannot detect a wrong value that occurs in the reference data, and considers a valid value invalid if it does not. 
Second, injected errors allow us to control their type and rate, but does not necessarily reflect the errors that occur in practice~\cite{jung2025realistic}. In addition, the LLM-based Semantic data accuracy DQM is currently limited to small tables or samples due to its runtime (see \autoref{sec:eval}).

\textbf{Challenge 1: Automation with LLMs.}
LLMs and other heuristics could automate DQ assessment further, as the Syntactic accuracy DQM already shows.
For example, LLMs could support the categorization (IG2) and parameterization (IG3) of DQMs, which shifts the effort from creating DQMs to validating them (IG5)~\cite{Glock_2026,Xie_2025,Chen_2026}. This may also apply to DQMs of category N, which currently cannot be automated because they require qualitative judgment.

\textbf{Challenge 2: Aggregation of DQM scores.}
Another challenge often emerging when investigating DQMs is the overwhelming complexity of large DQM reports. The interpretation of complex DQM reports could be simplified if DQ scores would be aggregated into a set of fewer high-level scores. Well-established (in literature as well as standards) high-level DQ dimensions such as accuracy, completeness, consistency, and timeliness are ideally suited for such aggregation. Hence, DQ measurement needs to be aligned to these concepts. Our proposed approach and reference implementation do precisely that as our DQMs are already mapped to the same vocabulary. Furthermore, as all DQMs are scaled into the same interval and have the same directionality (higher is better), aggregation could potentially be done by weighted averaging.
Yet the meaning of too aggressive aggregations over DQMs can quickly become difficult. One direction of reducing complexity while preserving the relevant tradeoffs in complex metric reports could take inspiration from the work on synthetic data generation for anonymization. Projecting the metrics onto a lower dimensional manifold can help to reduce the number of metrics substantially while preserving the relevant tradeoffs between metrics~\cite{nanevski2026evaluating}.

%% file: sections/7_conclusion.tex
\section{Conclusion}
\label{sec:conclusion}
The DQ measures in ISO/IEC 25024 and ISO/IEC 5259 define a comprehensive, yet abstract interface for assessing data quality in DBMS and ML applications.
In this work, we improved the applicability of these DQMs: we introduced a model of executable DQMs, classified the metrics of both standards into those that can be computed from data alone, those that require parameters, and those that require expert knowledge, and proposed five implementation guidelines that are not restricted to the ISO standards.
Putting our guidelines into practice, we introduce \ourpackage, a library of \numDQMs{} DQMs that learns the required parameters from clean reference data instead of  relying on expert knowledge and manual input. 
We demonstrate that the DQMs from \ourpackage detect injected errors, that their scores decrease  monotonically with diminishing DQ and ML performance, and that they scale linearly with the number of rows. 

In future work, we plan to extend \ourpackage to DQ dimensions and DQMs that are not covered so far, e.g., because they require more than one table or information beyond the data itself.
Moreover, we want to investigate how LLMs can support the categorization and parameterization of DQMs, including those that currently require qualitative judgment, and how the guidelines apply to DQMs beyond the ISO standards.

%% file: sections/8_artifacts.tex
\section{Artifacts}
The source code required to reproduce the results in this work is available in our public GitHub repository \href{https://github.com/calgo-lab/dqmeasure}{https://github.com/calgo-lab/dqmeasure}.
The repository contains the \ourpackage library, and instructions to reproduce the measurements we report, including access to all three datasets.